\documentclass[aps,preprint,a4paper,showpacs,showkeys,superscriptaddress,nofootinbib]{revtex4-1}

\usepackage{latexsym}
\usepackage{amsmath,amssymb}
\usepackage{graphicx}
\usepackage{subfigure}
\usepackage{xcolor}
\usepackage{cancel}
\usepackage{multirow,tabularx}

\newcolumntype{C}[1]{>{\centering\arraybackslash}p{#1}}
\usepackage{hyperref}     
\hypersetup{colorlinks,%
  citecolor=blue,%
  linkcolor=cyan,%
  pdftex}

\usepackage[titletoc]{appendix}
\usepackage{enumerate}

\begin{document}


\title{Quantum geodesics reflecting the internal structure of stars
 composed of shells}

\author{Sojeong Cheong}%
\email[]{jsquare@sogang.ac.kr}%
\affiliation{Department of Physics, Sogang University, Seoul, 04107,
  Republic of Korea}%
\affiliation{Center for Quantum Spacetime, Sogang University, Seoul, 04107, Republic of Korea}%

\author{Wontae Kim}%
\email[]{wtkim@sogang.ac.kr}%
\affiliation{Department of Physics, Sogang University, Seoul, 04107,
	Republic of Korea}%
\affiliation{Center for Quantum Spacetime, Sogang University, Seoul, 04107, Republic of Korea}%

\date{\today}

\begin{abstract}
    In general relativity, an external observer cannot distinguish distinct internal structures between two spherically symmetric stars that have the same total mass $M$.
    However, when quantum corrections are taken into account, the external metrics of the stars
    will receive quantum corrections depending on their internal structures.
    In this paper, we obtain the quantum-corrected metrics at linear order in curvature for two spherically symmetric shells
    characterized by different internal structures: one with an empty interior and the other with $N$ internal shells.
    The dependence on the internal structures in the corrected metrics tells us that geodesics on these backgrounds would be deformed according to the internal structures. We conduct numerical computations to find out the angle of geodesic precession and show that the presence of internal structures amplifies the precession angle reflecting the discrepancy between the radial and orbital periods within the geodesic orbit.
    The amount of the precession angle increases monotonically as the number of internal shells increases and it eventually converges to a certain value for $N \to \infty$.
\end{abstract}


\maketitle
\raggedbottom

\section{introduction}
\label{sec:introduction}
The significance of the connection between the conformal anomaly and the nonlocal effective action was emphasized by Deser, Duff,
and Isham~\cite{Deser:1976yx}.
Based on this work, Barvinsky, Vilkovisky and collaborators developed a useful effective action technique for computing the nonlocal quantum loop effects generated by gravitons or other massless fields~\cite{Barvinsky:1983vpp,Barvinsky:1985an,Barvinsky:1987uw,Barvinsky:1990uq,Avramidi:1990je,Donoghue:1994dn,Codello:2012kq,Donoghue:2014yha,Barvinsky:1993en,Barvinsky:1994hw}.
The technique involves an expansion in curvature and divides the terms of the effective action into local and nonlocal parts.
In the framework of effective field theory, tree-level calculations are integrated into the local part at low energies, as the curvature expansion corresponds to an energy expansion.
However, the nonlocal part resulting from one-loop calculations with higher-order curvature terms survives even at low energies.
(For a review, see~\cite{Buchbinder:1992rb,Donoghue:2017pgk}).
Equivalently, there is an alternative technique to explore the covariant nonlocal effective action, referred to as the \textit{non-linear completion}~\cite{Donoghue:2015nba,Donoghue:2015xla,El-Menoufi:2015cqw},
where the generalization of nonlocal operators from flat spacetime to curved spacetime has been studied.

There has been a wide variety of interest in using the effective action in various arenas, such as the Newtonian potential
~\cite{Donoghue:1994dn,Donoghue:1993eb,Bjerrum-Bohr:2002gqz},
black hole physics
~\cite{Calmet:2017qqa,Calmet:2018elv,Calmet:2021lny,Xiao:2021zly,Delgado:2022pcc,Battista:2023iyu},
gravitational waves
~\cite{Calmet:2016sba,Calmet:2017rxl,Calmet:2018rkj},
inflation
~\cite{Espriu:2005qn,Cabrer:2007xm},
and quantum hair
~\cite{Calmet:2021cip,Calmet:2022bpo,Calmet:2023gbw,Calmet:2023met}.
Particularly, it has been shown that the exterior metric of a star receives quantum effects that depend on the internal structure of the star~\cite{Calmet:2019eof,Calmet:2020tlj,Calmet:2021stu}.
These studies considered a star composed of nested solid spheres with constant mass densities and found that the quantum-corrected metric of the star depends on both the mass and the radius of the star's internal sphere.
This contrasts with general relativity, where the exterior vacuum solution to the field equations for a spherically symmetric object is uniquely determined by the Schwarzschild solution, irrespective of the internal structure of the object~\cite{1923rmp..book.....B}.

It is worth mentioning that several models for static black holes and static stars in
Refs.~\cite{Xiao:2021zly,Calmet:2017qqa,Calmet:2018elv,Calmet:2021lny,
Delgado:2022pcc,Calmet:2019eof,Calmet:2020tlj,Calmet:2021stu}
might exhibit certain limitations in the implementation of the Barvinsky-Vilkovisky curvature expansion.
In most physical problems including the collapse problem, spacetime is assumed to be asymptotically flat in the remote past~\cite{Barvinsky:1990up}.
In fact, the Barvinsky-Vilkovisky curvature expansion is applicable to problems with asymptotically flat boundary conditions in the Euclidean spacetime
in the remote past.
Therefore, the Barvinsky-Vilkovisky curvature expansion appears inapplicable to problems with permanently static sources.

We need to come up with a plausible scenario for the applicability of the Barvinsky-Vilkovisky approach to models with static objects, for example, a static spherical shell.
In the remote past, we first consider an adiabatically collapsing spherical shell with a very large radius, and its energy density is sufficiently small to assume a Minkowski spacetime approximation.
In late times, the collapsing shell can be approximately treated as a static one with a finite radius $R_0$ during a certain time interval,
analogous to a current static planet within our solar system.
On the other hand, during this period, we consider quantum corrections from a massless scalar field $\varphi$, distinct from the highly massive collapsing matter.
The quantum fluctuation of the scalar field around its vacuum expectation value, assumed to be zero,
is taken into account, and
this fluctuation is integrated out along the line of the Barvinsky-Vilkovisky formalism.
Then, the effective action for $S[g, \varphi]$
can be written as $S_{\rm eff}=\Gamma_{\rm eff}[g]$ without resorting to $\langle \varphi \rangle$.
In addition, we just consider the formal action $S_{\rm m}[g, f]$ for the spherical shell, where $f$ is a classical matter field for the shell totally decoupled from $\varphi$, but the static spherical source
is described by a stress tensor without an explicit form of the action.
The total action is assumed to be $S_{\rm tot} =S_{\rm EH}[g]+S_{\rm m}[g, f]+\Gamma_{\rm eff}[g]$, where the only quantum correction comes from $\varphi$ via the Barvinsky-Vilkovisky curvature expansion while the spherical shell is treated as a classical massive object.
Consequently, the Barvinsky-Vilkovisky curvature expansion may be applied to problems with static sources, interpreted approximately as a part of the collapsing model.

In this context, it would be interesting to study the internal structure of a star that consists of numerous internal shells.
In this model, various internal structures can be explored by changing the number of shells, while maintaining the total mass.
In this paper, we first consider two shells that have the same total mass and radius, but one has an empty interior, while the other contains a single shell with a smaller radius inside.
The quantum corrections to the exterior metrics of these shells are calculated at linear order in curvature, and then the calculations are extended to the case of $N$ internal shells.
To figure out the influence of internal structures, we investigate effective potentials and corresponding geodesics for a massive particle.
When the quantum corrections are taken into account, the particle would deviate from a perfect circular orbit.
In fact, the particle completing one radial period fails to return to its initial position due to a mismatch between the radial and orbital periods.
In this regard, we perform numerical calculations for a precession angle
quantifying the lack of the angle required for the particle to return to its initial position after one period of radial oscillation.
We show that the presence of internal structures amplifies the
precession angle:
the amount of the precession angle monotonically increases when the number of internal shells increases and eventually converges to a certain value as the number of internal shells approaches infinity.

The organization of this paper is as follows.
In Sec.~\ref{sec:corrections}, we will recapitulate the calculations of quantum gravitational corrections and then derive the leading order of correction to the exterior metric of $N$ internal shells.
In Sec.~\ref{sec:effective_potentials}, we will elucidate the quantum-corrected effective potentials and investigate their dependence on internal structures.
In Sec.~\ref{sec:geodesics}, we will study the geodesic orbit of the massive particle in order to obtain precession angles for various numbers of internal shells.
We will consider the range of the number of internal shells from $N=0$ to $N=20$, and extend our analysis to an infinite number of internal shells to demonstrate the effect of internal structures.
Finally, conclusion and discussion will be presented in Sec.~\ref{sec:conclusion}.

\section{Quantum corrections to shell geometries}
\label{sec:corrections}
We start with the total action as
    \begin{equation}\label{eq:total_effective_action}
        S_{\rm tot} =S_{\rm EH}[g]+S_{\rm m}[g, f]+\Gamma_{\rm eff}[g],
    \end{equation}
where $S_{\mathrm{EH}}[g]$ is the Einstein-Hilbert action and $S_{\mathrm{m}}[g, f]$ is the classical matter action for the shells.
The last term $\Gamma_{\mathrm{eff}}[g]$ is the quantum gravitational effective action at second order in curvature.
The effective action is divided into local and nonlocal parts, $\Gamma_\mathrm{L}$ and $\Gamma_\mathrm{NL}$, which are given by
\cite{Barvinsky:1983vpp,Barvinsky:1985an,Barvinsky:1987uw,Barvinsky:1990uq,Avramidi:1990je,Codello:2012kq,Donoghue:1994dn,Donoghue:2014yha,Calmet:2018elv}
    \begin{equation}\label{eq:local_action}
        \Gamma_\mathrm{L} = \int d^4 x \sqrt{-g} \left[ c_1(\mu) R^2 + c_2(\mu) R_{\mu\nu}R^{\mu\nu} + c_3(\mu) R_{\mu\nu\kappa\lambda}R^{\mu\nu\kappa\lambda} \right],
    \end{equation}
and
    \begin{equation}\label{eq:nonlocal_action}
        \Gamma_\mathrm{NL} = -\int d^4 x \sqrt{-g} \left[ \alpha R\ln\left(-\frac{\Box}{\mu^2}\right)R + \beta R_{\mu\nu}\ln\left(-\frac{\Box}{\mu^2}\right)R^{\mu\nu} + \gamma R_{\mu\nu\kappa\lambda}\ln\left(-\frac{\Box}{\mu^2}\right)R^{\mu\nu\kappa\lambda} \right]
    \end{equation}
with $\mu$ for an energy scale adopted for dimensional reasons
~\cite{Mukhanov:2007zz}.
Since the ultraviolet theory of quantum gravity has not been specified, the exact values of coefficients $c_1$, $c_2$, and $c_3$ in Eq.~\eqref{eq:local_action} remain undetermined
~\cite{Calmet:2019eof}.
However, the coefficients $\alpha$, $\beta$, and $\gamma$ in Eq.~\eqref{eq:nonlocal_action} can be obtained from
Ref.~\cite{Donoghue:2014yha}.
In particular, we will focus on a single scalar field, where the precise values of our interest are given by $\alpha=5(6\xi-1)^2/(11520\pi^2)$, $\beta=-2/(11520\pi^2)$, and $\gamma=2/(11520\pi^2)$ with $\xi$ representing the value of the nonminimal coupling for a scalar theory.
It should be noted that the last terms in Eqs.~\eqref{eq:local_action} and \eqref{eq:nonlocal_action} can be eliminated by using the local and nonlocal Gauss-Bonnet theorem at second order in curvature, along with the redefinition of coefficients as $\bar{c_1}=c_1-c_3$, $\bar{c_2}=c_2+4c_3$, $\bar{\alpha}=\alpha-\gamma$, and $\bar{\beta}=\beta+4\gamma$
~\cite{Calmet:2018elv}.

It is worth noting that the metric variation of the action includes terms that are one order lower than the order of the action.
The effective actions~\eqref{eq:local_action} and \eqref{eq:nonlocal_action} at second order in curvature generate both linear and higher-order terms.
If one were to consider the quadratic order of terms in the equations of motion, it would be necessary to take into account the cubic curvature terms of the effective action~\cite{Barvinsky:1990uq,Barvinsky:1993en}.
For simplicity, however, we will consider terms only up to linear order in curvature in the equations of motion for a consistent lowest-order perturbation.

Now, the equations of motion are obtained from the variation of the total action \eqref{eq:total_effective_action} with respect to the metric as
    \begin{equation}\label{eq:eom}
      G_{\mu\nu} + 16\pi G_\mathrm{N} (H^\mathrm{L}_{\mu\nu}+H^{\mathrm{NL}}_{\mu\nu}) = 8\pi G_\mathrm{N} T_{\mu\nu},
    \end{equation}
where $G_{\mu\nu}$ and $T_{\mu\nu}$ represent the Einstein tensor and the energy-momentum tensor, respectively.
Here, $H^\mathrm{L}_{\mu\nu}$ and $H^{\mathrm{NL}}_{\mu\nu}$, derived from the variations of Eqs.~\eqref{eq:local_action} and \eqref{eq:nonlocal_action}, are given by
    \begin{align}\label{eq:local_eom}
      H^\mathrm{L}_{\mu\nu} = &\bar{c_1} \left( 2g_{\mu\nu}\Box R - 2\nabla_\mu\nabla_\nu R \right) + \bar{c_2} \left( \Box R_{\mu\nu}  + \frac{1}{2}g_{\mu\nu}\Box R - \nabla_\alpha\nabla_\mu R^\alpha_\nu - \nabla_\alpha\nabla_\nu R^\alpha_\mu \right)
    \end{align}
and
    \begin{align}\label{eq:nonlocal_eom}
      H^{\mathrm{NL}}_{\mu\nu} = &-2\bar{\alpha} \left( g_{\mu\nu}\Box - \nabla_\mu\nabla_\nu \right)\ln\left(-\frac{\Box}{\mu^2}\right)R\nonumber\\
          &- \bar{\beta} \left( \delta^\alpha_\mu g_{\nu\beta}\Box + g_{\mu\nu}\nabla^\alpha\nabla_\beta - \delta^\alpha_\mu\nabla_\beta\nabla_\nu - \delta^\alpha_\nu\nabla_\beta\nabla_\mu \right)\ln\left(-\frac{\Box}{\mu^2}\right)R^\beta_\alpha.
    \end{align}
Note that the metric variation of the logarithmic operator in the nonlocal effective action~\eqref{eq:nonlocal_action} produces terms higher than quadratic
in curvature, as demonstrated in Refs.~\cite{Donoghue:2014yha,Calmet:2018elv,Donoghue:2015nba}.

The energy-momentum tensor for a static and spherically symmetric dust shell with mass $M$ is described by
    \begin{equation}\label{eq:energy_momentum_tensor}
      T_{\mu\nu} = \rho(r) u_\mu u_\nu,
    \end{equation}
where $u_\mu$ is the four-velocity of the shell, satisfying the normalization
condition of $u^\mu u_\mu =-1$, and the energy density is chosen as $\rho(r) = \sigma_0 \delta(r-R_0)$.
Note that $\sigma_0$ denotes the constant mass density of the shell, and it satisfies $M = 4 \pi R_0^2 \sigma_0$, where $R_0$ is the radius of the shell.

Without the effective actions, the equations of motion reduce to
    \begin{equation}\label{eq:no_eff_action_eom}
        G^{(0)}_{\mu\nu} - 8\pi G_\mathrm{N} T_{\mu\nu} = 0.
    \end{equation}
The solution to Eq.~\eqref{eq:no_eff_action_eom} is the Schwarzschild metric given by
    \begin{align}\label{eq:Schwarzschild_metric}
      ds^2 &= g^{(0)}_{\mu\nu} dx^\mu dx^\nu\nonumber\\
           &= - \left( 1 - \frac{2G_\mathrm{N}M}{r} \right) dt^2 + \left( 1 - \frac{2G_\mathrm{N}M}{r} \right)^{-1} dr^2 + r^2 d\Omega^2,
    \end{align}
which will be regarded as the background metric.
To solve the quantum-corrected equations of motion~\eqref{eq:eom}, we now consider the metric perturbation of
    \begin{equation}\label{eq:metric_perturbation}
      g_{\mu\nu} = g^{(0)}_{\mu\nu} + h_{\mu\nu},
    \end{equation}
where the perturbation $h_{\mu\nu}$ is assumed to be $\mathcal{O}(G_\mathrm{N})$,
in particular,
    \begin{equation}
        h_{\mu\nu}=G_\mathrm{N} h_{1 \mu\nu} + G_\mathrm{N}^2 h_{2 \mu\nu}
    \end{equation}
up to the quadratic order of $G_\mathrm{N}$.
Inserting Eq.~\eqref{eq:metric_perturbation} into Eq.~\eqref{eq:eom}, we obtain:
    \begin{align}
      \text{$G_\mathrm{N}$ order:} \quad &G^{(h)}_{\mu\nu}\left[h_1\right] = 0 \label{eq:classical_eom},\\
      \text{$G_\mathrm{N}^2$ order:} \quad &G_{\mu\nu}^{(h)}\left[h_2\right] + 16\pi G_\mathrm{N} \left\{ \bar{c_1}[2\eta_{\mu\nu}\bar{\square} R^{(0)} - 2\bar{\nabla}_\mu \bar{\nabla}_\nu R^{(0)}] \right. \nonumber\\
    &+ \bar{c_2}[\bar{\square} R_{\mu\nu}^{(0)} + \frac{1}{2}\eta_{\mu\nu}\bar{\square} R^{(0)} - \bar{\nabla}_\alpha \bar{\nabla}_\mu R^{\alpha(0)}_\nu - \bar{\nabla}_\alpha \bar{\nabla}_\nu R^{\alpha(0)}_\mu] \nonumber\\
    &-2\bar{\alpha}[\eta_{\mu\nu}\bar{\square} - \bar{\nabla}_\mu \bar{\nabla}_\nu] \ln\left(-\frac{\bar{\square}}{\mu^2}\right)R^{(0)} \nonumber\\
    &\left. - \bar{\beta}[\delta^\alpha_\mu \eta_{\nu\beta}\bar{\square} + \eta_{\mu\nu}\bar{\nabla}^\alpha \bar{\nabla}_\beta - \delta^\alpha_\mu \bar{\nabla}_\beta \bar{\nabla}_\nu - \delta^\alpha_\nu \bar{\nabla}_\beta \bar{\nabla}_\mu]\ln\left(-\frac{\bar{\square}}{\mu^2}\right)R^{\beta(0)}_\alpha \right\} = 0 \label{eq:perturbed_eom},
    \end{align}
where the linearized Einstein tensor $G^{(h)}_{\mu\nu}$ in Eqs.~\eqref{eq:classical_eom} and \eqref{eq:perturbed_eom} is given by
    \begin{equation}\label{eq:linearized_Einstein_tensor}
      2G^{(h)}_{\mu\nu} = -\bar{\square} h_{\mu\nu} + \eta_{\mu\nu}\bar{\square} h - \bar{\nabla}_\mu\bar{\nabla}_\nu h + \bar{\nabla}^\alpha\bar{\nabla}_\mu h_{\nu\alpha} + \bar{\nabla}^\alpha\bar{\nabla}_\nu h_{\mu\alpha} - \eta_{\mu\nu}\bar{\nabla}^\alpha\bar{\nabla}^\beta h_{\alpha\beta}
    \end{equation}
and $h = h_{\mu}^\mu$, and both $\bar{\square}$ and $\bar{\nabla}_\mu$ are defined with respect to the flat metric $\eta_{\mu\nu}$.
Note that the orders of $R^{(0)}$ and $R^{\alpha (0)}_\beta$ are linear in $G_\mathrm{N}$; hence, the logarithmic operator in Eq.~\eqref{eq:eom} reduces to $\ln(-\bar{\square}/\mu^2)$, which is the flat spacetime approximation of the nonlocal logarithmic operator
\cite{Calmet:2017qqa}.
As for the $G_\mathrm{N}$ order, in the absence of the zeroth order of terms in $H^{\mathrm{L}(0)}_{\mu\nu}$ and $H^{\mathrm{NL}(0)}_{\mu\nu}$,
Eq.~\eqref{eq:classical_eom} yields $h_{1 tt} = C_1/r + C_2$ and $h_{1 rr} = C_1/r$ for the integration constants $C_1$ and $C_2$, where we fixed the gauge as $h_{\theta\theta} = h_{\phi\phi} = 0$.
To require the asymptotic flatness and to recover the classical weak field limit with ADM mass as $r\to\infty$, $C_1$ and $C_2$ are chosen to be zero.
Thus, the perturbed metric can be written up to the second order of $G_\mathrm{N}$ as $h_{\mu\nu} = G_\mathrm{N}^2 h_{2 \mu\nu}$.

In order to solve Eq.~\eqref{eq:perturbed_eom}, we will use
$\ln(-\bar{\square}/\mu^2)$ acting on
spherically symmetric time-independent functions,
which is calculated as~\cite{Calmet:2019eof}
    \begin{align}\label{eq:log_box_general}
      \ln \left(-\frac{\bar{\square}}{\mu^2}\right) f(r) = &\frac{1}{\pi r} \int^\infty_0 dr' f(r') r' \lim_{\epsilon \rightarrow 0^+} \left[ \frac{\gamma_E + \ln(\mu R_+) + i\phi_+}{\epsilon + i(r+r')} + \frac{\gamma_E + \ln(\mu R_+) - i\phi_+}{\epsilon - i(r+r')} \right. \nonumber\\
      &\left. - \frac{\gamma_E + \ln(\mu R_-) + i\phi_-}{\epsilon + i(r-r')} - \frac{\gamma_E + \ln(\mu R_-) - i\phi_-}{\epsilon - i(r-r')} \right],
    \end{align}
where $R_\pm = \sqrt{\epsilon^2+(r\pm r')^2}$, $\phi_\pm = \arctan [(r\pm r')/\epsilon]$.
If Ricci is zero everywhere, both the local part $H^{\mathrm{L}(0)}_{\mu\nu}$ and the nonlocal part $H^{\mathrm{NL}(0)}_{\mu\nu}$ must vanish while the presence of a single shell at $r=R_0$ results in a nonvanishing Ricci scalar and Ricci tensor such as $R^{(0)}=(2G_\mathrm{N} M/R_0^2)\delta(r-R_0)$ and $R^{\beta (0)}_\alpha=-(2G_\mathrm{N} M/R_0^2)\delta(r-R_0)(\delta^\beta_0\delta^0_\alpha-\delta^\beta_\alpha/2)+\mathcal{O}(G_\mathrm{N}^2)$, respectively.
In the region $R_0<r<\infty$, the local part $H^{\mathrm{L}(0)}_{\mu\nu}=0$ again.
However, $H^{\mathrm{NL}(0)}_{\mu\nu}$ possesses the nonlocal log operator covering
$0<r<\infty$ so that
the non-vanishing $H^{\mathrm{NL}(0)}_{\mu\nu}$ can be obtained owing to contributions from the delta function.
When $f(r')=0$ around $r$, Eq.~\eqref{eq:log_box_general} reduces to
    \begin{equation}\label{eq:log_box_general_result}
      \ln \left(-\frac{\bar{\square}}{\mu^2}\right) f(r) = \frac{1}{r} \int^{\infty}_0 \left( \frac{r'}{r+r'} - \frac{r'}{|r-r'|} \right) f(r') dr'
    \end{equation}
by taking the limit $\epsilon \rightarrow 0^+$.
If $f(r)=\delta(r-R_0)$, Eq.~\eqref{eq:log_box_general_result} is computed as
    \begin{equation}\label{eq:log_box_delta}
      \ln \left(-\frac{\bar{\square}}{\mu^2}\right) \delta(r-R_0) = \frac{R_0}{r} \left( \frac{1}{r+R_0} - \frac{1}{r-R_0} \right),
    \end{equation}
where $r>R_0$.
Hence, the corrections for the shell with an empty interior are obtained as
    \begin{align}
      h_{tt} &= \frac{128\pi \bar{\gamma} G_\mathrm{N}^2 M}{r(r^2-R_0^2)} + \frac{C_3}{r} + C_4 + \mathcal{O}(G_\mathrm{N}^3),\label{eq:corrected_metric_oneshell_tt}\\
      h_{rr} &= - \frac{128\pi \bar{\alpha} G_\mathrm{N}^2 M (3r^2-R_0^2)}{r(r^2-R_0^2)^2} + \frac{C_3}{r} + \mathcal{O}(G_\mathrm{N}^3),\label{eq:corrected_metric_oneshell_rr}
    \end{align}
where we
redefined $\bar{\gamma}=\bar{\alpha}+\bar{\beta}=\alpha+\beta+3\gamma$.
The integration constant $C_3$ is determined to be zero to recover the classical weak field limit with ADM mass as $r \rightarrow \infty$, and the other constant $C_4$ is also set to zero in order to ensure the asymptotic flatness of spacetime.

We can extend the corrections for the empty interior to configurations with $N$ internal shells.
Let us first consider a scenario with the shell containing a single internal shell of $N=1$.
The radius $R_0$ of the outermost shell is the same as in the case of the empty interior shell, while the radius of the internal shell is denoted by $R_1$, satisfying $R_0>R_1$.
The masses of the outermost shell and the internal shell are represented by $M_0$ and $M_1$, respectively, with the total mass maintained such that $M_0+M_1=M$.
The energy-momentum tensor is described by the energy density $\rho(r) = \sigma_0 \delta(r-R_0) + \sigma_1 \delta(r-R_1)$, where $M_0 = 4 \pi R_0^2 \sigma_0$ and $M_1 = 4 \pi R_1^2 \sigma_1$ with $\sigma_{0}$ and $\sigma_{1}$ being constant mass densities.
Then, we obtain the quantum-corrected metric components for $N=1$:
    \begin{align}
      h_{tt}^{(1)} &= 128\pi \bar{\gamma} G_\mathrm{N}^2 \left[\frac{M_0}{r(r^2-R_0^2)} + \frac{M_1}{r(r^2-R_1^2)} \right] + \frac{C_5}{r} + C_6 + \mathcal{O}(G_\mathrm{N}^3),\label{eq:corrected_metric_twoshells_tt}\\
      h_{rr}^{(1)} &= - 128\pi \bar{\alpha} G_\mathrm{N}^2 \left[\frac{M_0 (3r^2-R_0^2)}{r(r^2-R_0^2)^2} + \frac{M_1 (3r^2-R_1^2)}{r(r^2-R_1^2)^2} \right] + \frac{C_5}{r} + \mathcal{O}(G_\mathrm{N}^3),\label{eq:corrected_metric_twoshells_rr}
    \end{align}
where the integration constants $C_5$ and $C_6$ are determined to be zero in virtue of the aforementioned requirements.
Note that Eqs.~\eqref{eq:corrected_metric_twoshells_tt} and \eqref{eq:corrected_metric_twoshells_rr} reduce to Eqs.~\eqref{eq:corrected_metric_oneshell_tt} and \eqref{eq:corrected_metric_oneshell_rr} in the limit $M_1 \rightarrow 0$.

For a shell comprising $N$ internal shells, the quantum-corrected metric components can easily be obtained as
    \begin{align}
      h_{tt}^{(N)} &= 128\pi \bar{\gamma} G_\mathrm{N}^2 \sum^{N}_{n=0} \frac{M_n}{r(r^2-R_n^2)} + \mathcal{O}(G_\mathrm{N}^3),\label{eq:corrected_metric_Nshell_tt}\\
      h_{rr}^{(N)} &= - 128\pi \bar{\alpha} G_\mathrm{N}^2 \sum^{N}_{n=0} \frac{M_n (3r^2-R_n^2)}{r(r^2-R_n^2)^2} + \mathcal{O}(G_\mathrm{N}^3),\label{eq:corrected_metric_Nshell_rr}
    \end{align}
where $R_0>R_1>\dots>R_N$, $M_n = 4 \pi R_n^2 \sigma_n$ for $n=0,1,\dots,N$, and $\Sigma^N_{n=0} M_n = M$.
Since the corrections are infinite as $r$ approaches $R_n$ due to the discontinuity of the matter source~\cite{Calmet:2019eof}, we will consider a region far from the outermost shell, where $r \gg R_0$.
Hence, the quantum-corrected metric components for $N$ internal shells can be expressed as
    \begin{align}
      h_{tt}^{(N)} &= \frac{128\pi \bar{\gamma} G_\mathrm{N}^2 M}{r^3} + \frac{128\pi \bar{\gamma} G_\mathrm{N}^2}{r^3} \sum^{N}_{n=0}\left[ \frac{M_n R_n^2}{r^2} + \mathcal{O}\left(R_0/r\right)^4 \right] + \mathcal{O}(G_\mathrm{N}^3),\label{eq:corrected_metric_Taylor_Nshell_tt}\\
      h_{rr}^{(N)} &= - \frac{384\pi \bar{\alpha} G_\mathrm{N}^2 M}{r^3} - \frac{640\pi \bar{\alpha} G_\mathrm{N}^2}{r^3} \sum^{N}_{n=0}\left[ \frac{M_n R_n^2}{r^2} + \mathcal{O}\left(R_0/r\right)^4 \right] + \mathcal{O}(G_\mathrm{N}^3)\label{eq:corrected_metric_Taylor_Nshell_rr}
    \end{align}
which reduce to the corrections for the shell with an empty interior when $N=0$.
It is worth noting that the corrections proportional to  $r^{-5}$ depend on the composition of the shells and so they are more sensitive to larger or more massive internal shells.

\section{Quantum-corrected effective potentials}
\label{sec:effective_potentials}
Let us now consider the equatorial orbit for a massive particle, which is defined on the plane where $\theta = \pi/2$.
From Eq.~\eqref{eq:metric_perturbation} together with
Eqs.~\eqref{eq:corrected_metric_Taylor_Nshell_tt} and \eqref{eq:corrected_metric_Taylor_Nshell_rr},
the metric of the shell composed of $N$ internal shells for $r\gg R_0$ can be
written as
    \begin{equation}\label{eq:metric_oneshell}
      ds^2 = -f(r)dt^2 + g(r)dr^2 + r^2d\phi^2,
    \end{equation}
where
    \begin{align}
        f(r) = &1 - \frac{2G_\mathrm{N} M}{r} - \frac{128\pi \bar{\gamma} G_\mathrm{N}^2 M}{r^3} - \frac{128\pi \bar{\gamma} G_\mathrm{N}^2}{r^3} \sum^{N}_{n=0} \left[ \frac{M_n R_n^2}{r^2} + \mathcal{O}\left(R_0/r\right)^4 \right] + \mathcal{O}(G_\mathrm{N}^3),\label{eq:metric_function_for_t_Nshell}\\
        g(r) = &\left( 1 - \frac{2G_\mathrm{N} M}{r} \right)^{-1} - \frac{384\pi \bar{\alpha} G_\mathrm{N}^2 M}{r^3} - \frac{640\pi \bar{\alpha} G_\mathrm{N}^2}{r^3} \sum^{N}_{n=0} \left[ \frac{M_n R_n^2}{r^2} + \mathcal{O}\left(R_0/r\right)^4 \right] + \mathcal{O}(G_\mathrm{N}^3).\label{eq:metric_function_for_r_Nshell}
    \end{align}
Since the metric functions are independent of $t$ and $\phi$, the conserved quantities for the particle can be defined by
    \begin{align}
        E &\equiv -u_\mu \xi^\mu_{(t)} = f(r) \dot{t},\label{eq:particle_energy}\\
        L &\equiv u_\mu \xi^\mu_{(\phi)} = r^2 \dot{\phi},\label{eq:particle_angular_momentum}
    \end{align}
where $E$ and $L$ are the conserved energy and angular momentum per unit mass of the particle, respectively.
The overdot denotes a derivative with respect to the affine parameter $\tau$.
In Eqs.~\eqref{eq:particle_energy} and \eqref{eq:particle_angular_momentum}, $\xi^\mu_{(t)}$ and $\xi^\mu_{(\phi)}$ are the Killing vectors corresponding to the time and azimuthal coordinates, respectively.
Then, the normalization condition $u_\mu u^\mu=-1$ leads to the following equation:
    \begin{equation}\label{eq:energy_equation}
      -\frac{E^2}{f(r)} + g(r)\dot{r}^2 + \frac{L^2}{r^2} = -1.
    \end{equation}

Next, the effective potential is defined as~\cite{Hartle:2003yu}
    \begin{equation}\label{eq:effective_potential_def}
      \frac{E^2-1}{2} = \frac{1}{2}\dot{r}^2 + V^{(N)}_{\mathrm{eff}}(r)
    \end{equation}
which becomes
    \begin{equation}\label{eq:effective_potential_definition}
      V^{(N)}_{\mathrm{eff}}(r) = \frac{E^2-1}{2} + \frac{1}{2g(r)}\left( -\frac{E^2}{f(r)} + \frac{L^2}{r^2} +1 \right)
    \end{equation}
after eliminating $\dot{r}$ in Eq.~\eqref{eq:effective_potential_def} by the use of Eq.~\eqref{eq:energy_equation}.
Upon substituting Eqs.~\eqref{eq:metric_function_for_t_Nshell} and \eqref{eq:metric_function_for_r_Nshell} into Eq.~\eqref{eq:effective_potential_definition},
we get the effective potential as
    \begin{align}\label{eq:effective_potential_far_Nshell}
        V^{(N)}_{\mathrm{eff}}(r) = &-\frac{G_\mathrm{N} M}{r} + \frac{L^2}{2r^2} - \frac{G_\mathrm{N} M}{r^3} \left[ L^2 + 64\pi G_\mathrm{N} \left\{3\bar{\alpha}(E^2-1) + E^2\bar{\gamma}\right\} \right] \nonumber\\
        &- \frac{64\pi G_\mathrm{N}^2}{r^3}\! \left[ \sum^{N}_{n=0} \frac{M_n R_n^2}{r^2} \left\{5\bar{\alpha}(E^2\!-\!1) \!+\! E^2\bar{\gamma}\right\} \!-\! 3\bar{\alpha}ML^2 \!+\! \mathcal{O}\left(R_0/r\right)^4 \right]\!+\! \mathcal{O}(G_\mathrm{N}^3),
    \end{align}
whereas the classical effective potential is defined as
    \begin{equation}\label{eq:effective_potential_classical}
        V^{(\mathrm{cl})}_{\mathrm{eff}}(r) = -\frac{G_\mathrm{N} M}{r} + \frac{L^2}{2r^2} - \frac{G_\mathrm{N} M L^2}{r^3},
    \end{equation}
where the local minimum of the classical potential occurs at $r_0 = L^2(1 + \sqrt{1 \!-\! 12G_\mathrm{N}^2 M^2/L^2})/(2G_\mathrm{N}M)$, which is the radius of the stable circular motion.
Obviously, the quantum corrections affect the classical circular orbit so that the orbit does not maintain its radius $r_0$.
Since the orbital shape would be deformed due to the $r^{-5}$ corrections depending on the internal structure, a distant observer can distinguish the internal structures of shells through the orbit of the particle.
In this regard, quantum deformations of the geodesic orbit of the particle
will be numerically investigated for various numbers of internal shells
in the next section.

\section{Numerical analysis of geodesic deviations}
\label{sec:geodesics}

In this section, we numerically solve the geodesic equations \eqref{eq:particle_energy}-\eqref{eq:energy_equation} based on the quantum-corrected metrics.
For a stable circular orbit of a massive particle at radius $r_0$ in the absence of quantum corrections, we set the values of the energy $E$ and angular momentum $L$.
Henceforth, the values of $E$, $L$, $M$, $R_0$, and the initial radial position of the particle $r_0$ are treated as fixed constants throughout our calculations, in order to compare the quantum-corrected orbital motion with the classical circular orbit.
To conduct numerical calculations, we employ a shell which contains $N$ equidistant additional internal shells.
In this model, the shells are assumed to be distributed uniformly within a fixed range between $2G_\mathrm{N}M$ and $R_0$, each with an equal mass while maintaining the total mass $M$. So,
the mass and radius of an $n$-th shell are determined by
    \begin{align}
        M_n &= \frac{M}{N+1},\label{eq:Mn}\\
        R_n &= R_0 - n \left( \frac{R_0-2G_\mathrm{N}M}{N+1} \right),\label{eq:Rn}
    \end{align}
where the denominators account for the total number of shells, including not only the $N$ internal shells but also the outermost shell.
Then, plugging Eqs.~\eqref{eq:Mn} and \eqref{eq:Rn} into Eqs.~\eqref{eq:corrected_metric_Taylor_Nshell_tt} and \eqref{eq:corrected_metric_Taylor_Nshell_rr}, we obtain the quantum-corrected metric functions in Eq.~\eqref{eq:metric_perturbation} as
    \begin{align}
      f(r) = &1 - \frac{2G_\mathrm{N} M}{r} - \frac{128\pi \bar{\gamma} G_\mathrm{N}^2 M}{r^3}\nonumber\\
      &- \frac{128\pi \bar{\gamma} G_\mathrm{N}^2}{r^3} \left[ \frac{MR_0^2}{r^2}
      \!-\! \frac{MR_0^2N}{r^2(N+1)}
      \!+\! \frac{MR_0^2N(2N+1)}{6r^2(N+1)^2} \!+\! \mathcal{O}\left(R_0/r\right)^4 \right] +\mathcal{O}(G_\mathrm{N}^3),\label{eq:corrected_metric_Taylor_equidistant_tt}\\
      g(r) = &\left( 1 - \frac{2G_\mathrm{N} M}{r} \right)^{-1} - \frac{384\pi \bar{\alpha} G_\mathrm{N}^2 M}{r^3}\nonumber\\
      &- \frac{640\pi \bar{\alpha} G_\mathrm{N}^2}{r^3} \left[ \frac{MR_0^2}{r^2} \!-\! \frac{MR_0^2N}{r^2(N+1)}
      \!+\! \frac{MR_0^2N(2N+1)}{6r^2(N+1)^2} \!+\! \mathcal{O}\left(R_0/r\right)^4 \right]
      +\mathcal{O}(G_\mathrm{N}^3)\label{eq:corrected_metric_Taylor_equidistant_rr}
    \end{align}
for $r\gg R_0$.

\begin{figure}[t]
\centering
\subfigure[~No internal shell~($N=0$)]{\includegraphics[width=0.48\textwidth]{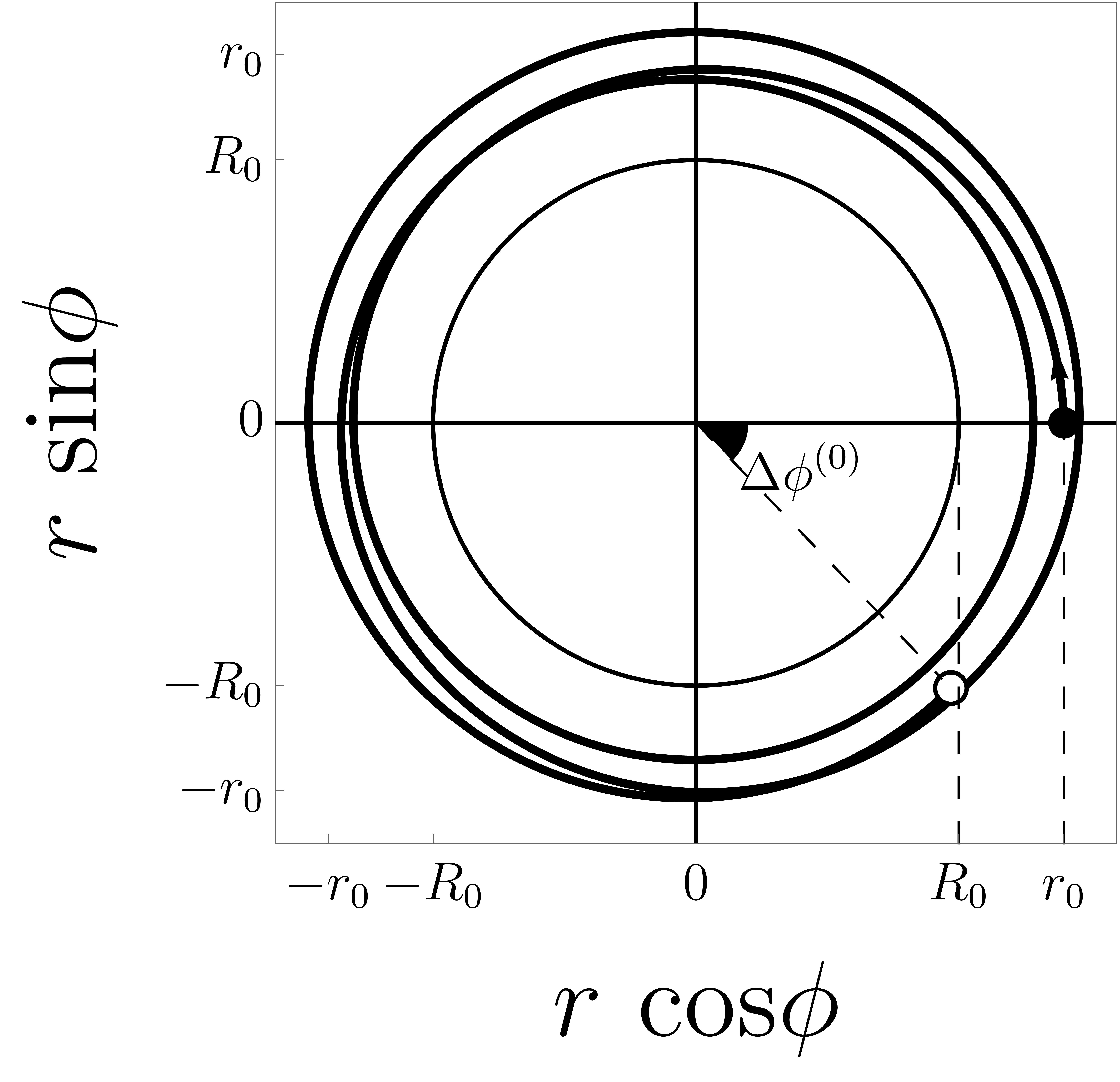}\label{fig:0shell_corrected}}
\subfigure[~A single internal shell~($N=1$)]{\includegraphics[width=0.48\textwidth]{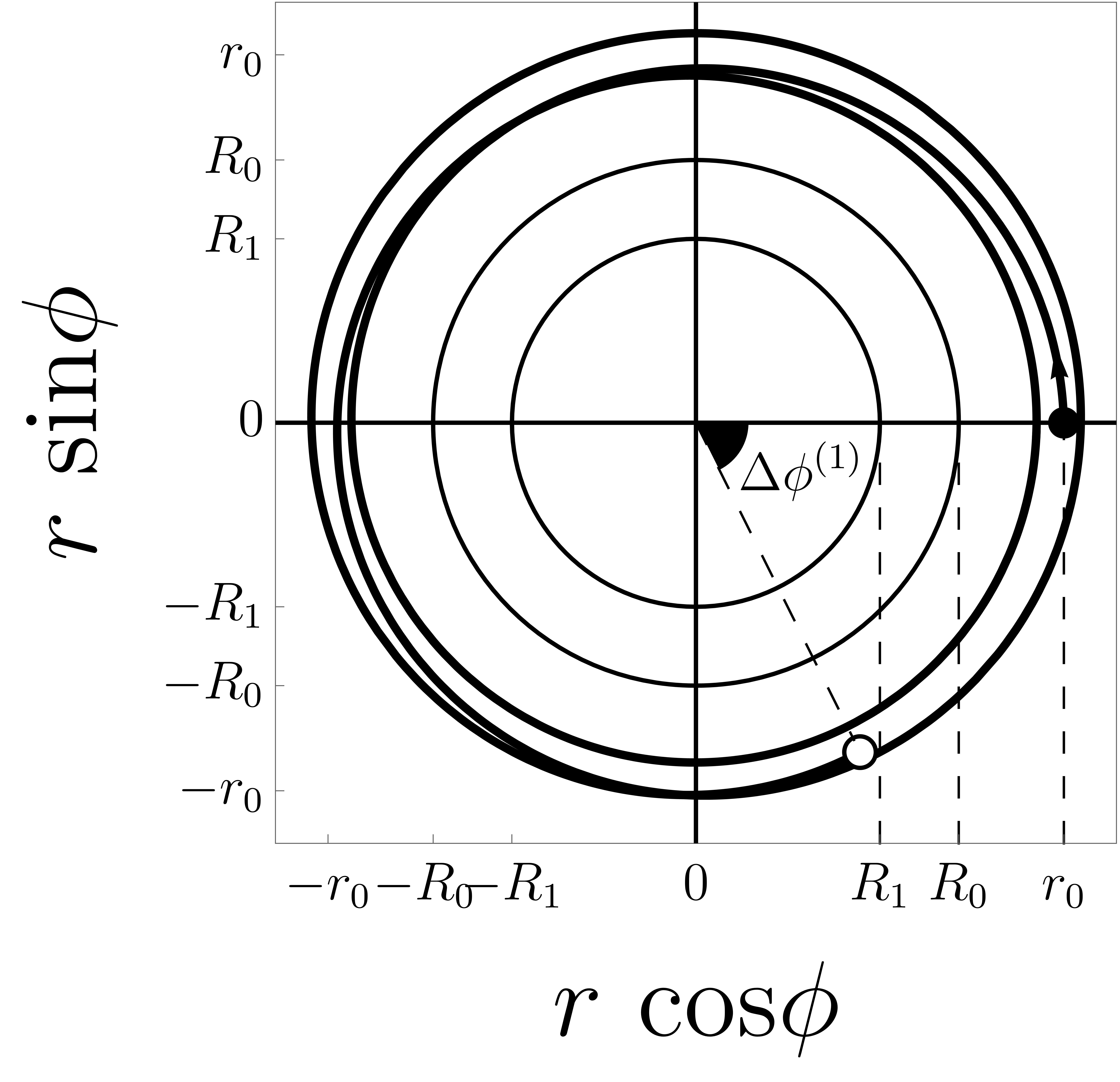}\label{fig:1shell_corrected}}
\caption{The quantum-corrected geodesics for $N=0$ and $N=1$ are plotted as solid curves in Figs.~\ref{fig:0shell_corrected} and \ref{fig:1shell_corrected}, respectively.
Each solid curve represents the trajectory of a massive particle completing one period of radial oscillation, and the thin circles depict the mass shells.
The starting and end points of the particle within the radial period are presented as the black dot and the white dot, respectively.
The constants are fixed as $E=5/(2\sqrt{7})$, $L=7/2$, $M=1$, $M_0=M_1=1/2$, $R_0=5$, and $R_1=7/2$ with $G_\mathrm{N}=1$.
}
\end{figure}

In Figs.~\ref{fig:0shell_corrected} and \ref{fig:1shell_corrected}, we present the geodesic orbits of massive particles for $N=0$ and $N=1$, respectively.
The quantum-corrected trajectories deviate from the classical trajectories.
Of course, the radial distance $r(\tau)$ of the quantum-corrected trajectory can still be confined to a finite range such as $r_{\mathrm{min}}<r(\tau)<r_{\mathrm{max}}$.
Here, $r_{\mathrm{min}}$ and $r_{\mathrm{max}}$ denote the periastron and apastron radii, respectively, representing the turning points that can be derived from Eq.~\eqref{eq:effective_potential_definition}.
For $N=0$, the numerical values of those radii are found to be $r_{\mathrm{min}} \approx 6.402$ and $r_{\mathrm{max}} \approx 7.436$, whereas for $N=1$, the values are approximately $r_{\mathrm{min}} \approx 6.457$ and $r_{\mathrm{max}} \approx 7.412$.
This fact indicates that for given $E$ and $L$ the effective potential \eqref{eq:effective_potential_far_Nshell} for $N=1$ is shifted upward compared to that of $N=0$ and thus, $r_{\mathrm{min}}$ gets larger and $r_{\mathrm{max}}$ gets smaller.

The quantum corrections modify the period of radial oscillation, and so we define $\Delta \phi^{(N)}$ as a precession angle to quantify the discrepancy between the period of radial oscillation and the orbital period around the shell comprising $N$ internal shells.
Note that the period of radial oscillation is defined as the time required to complete an orbit starting from $r_0$, reaching the turning points at $r_\mathrm{min}$ and $r_\mathrm{max}$ each once, and subsequently returning to $r_0$.
In Fig.~\ref{fig:0shell_corrected}, the numerical value of $\Delta \phi^{(0)}$ is calculated as $\Delta \phi^{(0)} \approx 0.805$, whereas in Fig.~\ref{fig:1shell_corrected}, $\Delta \phi^{(1)} \approx 1.108$.
These values reveal that the presence of the internal structure amplifies the deviation between radial and orbital periods.

\begin{figure}[t]
\centering
\includegraphics[width=0.6\textwidth]{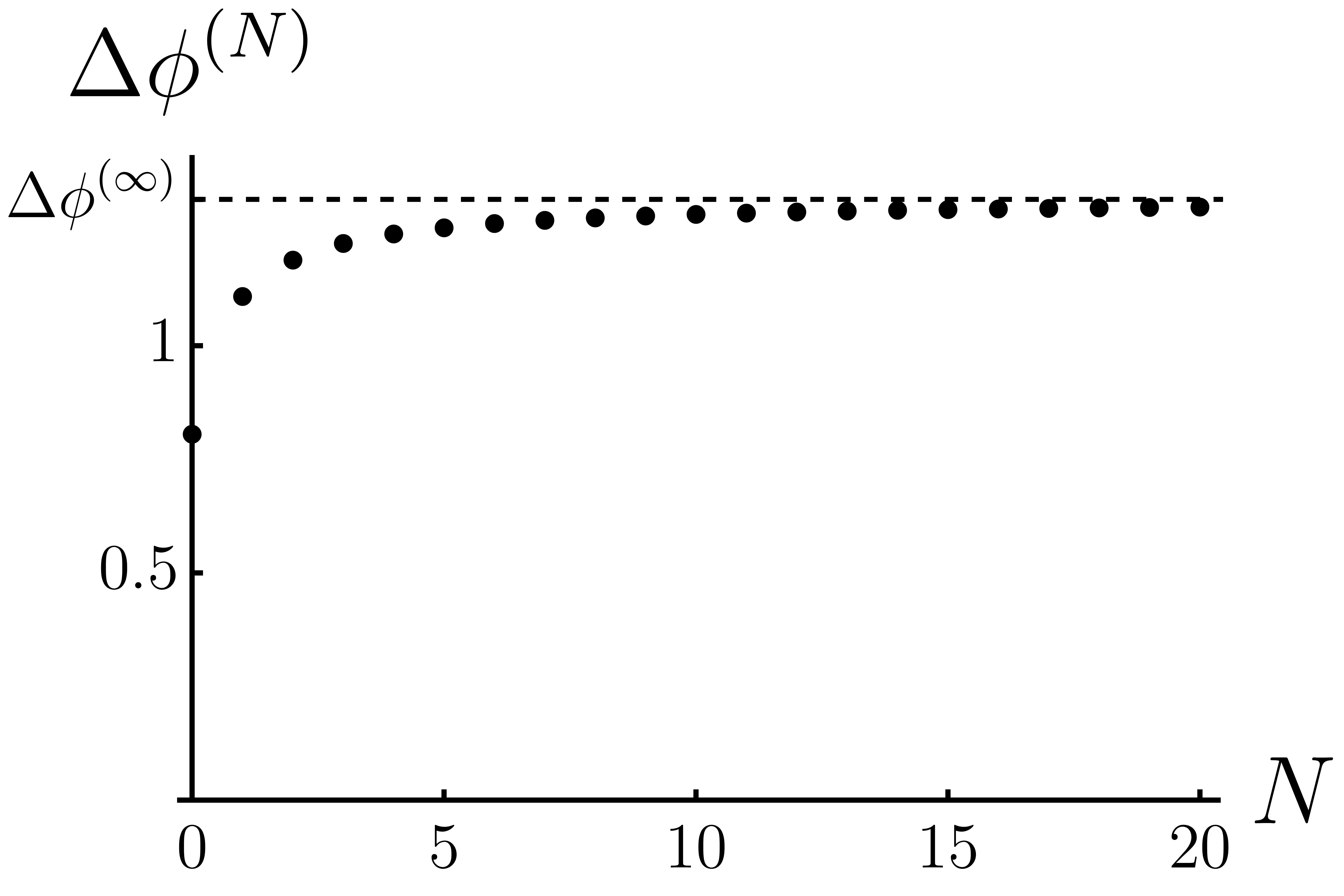}
\caption{The precession angles are presented for $N=0$ to $N=20$.
The $N$ shells are distributed equidistantly within the range from $2G_\mathrm{N}M$ to $R_0$, with an equal mass of each shell while the total mass is maintained as $M=1$.
The numerical values of the precession angle are $\Delta \phi^{(0)} \approx 0.805$ and $\Delta \phi^{(20)} \approx 1.305$ for $N=0$ and $N=20$, respectively, and extremely $\Delta \phi^{(\infty)} \approx 1.406$ for $N \rightarrow \infty$.
}
\label{fig:Nshell}
\end{figure}

We further perform the numerical calculations for various $N$.
The calculations show that $\Delta \phi^{(N)}$ monotonically increases as the number of shells increases; for example, this feature from $N=0$ to $N=20$ is illustrated in Fig.~\ref{fig:Nshell}.
In the limit $N \rightarrow \infty$ of Eqs.~\eqref{eq:corrected_metric_Taylor_equidistant_tt} and \eqref{eq:corrected_metric_Taylor_equidistant_rr}, the quantum-corrected metric components for an infinite number of shells are easily obtained as
    \begin{align}
      h_{tt}^{(\infty)} &= \frac{128\pi \bar{\gamma} G_\mathrm{N}^2 M}{r^3} + \frac{128\pi \bar{\gamma} G_\mathrm{N}^2}{3r^3}\left[ \frac{MR_0^2}{r^2} + \mathcal{O}\left(R_0/r\right)^4 \right] + \mathcal{O}(G_\mathrm{N}^3),\label{eq:corrected_metric_Taylor_equidistant_Ninfty_tt}\\
      h_{rr}^{(\infty)} &= - \frac{384\pi \bar{\alpha} G_\mathrm{N}^2 M}{r^3} - \frac{640\pi \bar{\alpha} G_\mathrm{N}^2}{3r^3} \left[ \frac{MR_0^2}{r^2} + \mathcal{O}\left(R_0/r\right)^4 \right] + \mathcal{O}(G_\mathrm{N}^3).\label{eq:corrected_metric_Taylor_equidistant_Ninfty_rr}
    \end{align}
Using Eqs.~\eqref{eq:corrected_metric_Taylor_equidistant_Ninfty_tt} and \eqref{eq:corrected_metric_Taylor_equidistant_Ninfty_rr}, we find that $\Delta \phi^{(\infty)} \approx 1.406$, which is depicted as a dashed horizontal line in Fig.~\ref{fig:Nshell}.
The precession angle converges to the finite value when the number of shells goes to infinity.

\section{conclusion and discussion}
\label{sec:conclusion}
We investigated the quantum corrections to the metrics of spherical shells with various internal structures at linear order in curvature and probed their implications on geodesic motion.
Firstly, we derived the quantum-corrected metric for the shell with an empty interior from the quantum gravitational effective action.
We then generalized our analysis to the shell encompassing $N$ internal shells.
The mass distribution for the shells is modeled as a linear combination of delta functions with constant surface densities.
To circumvent the divergences occurring as $r$ approaches $R_n$, we expanded the corrected metric in the far region of $r\gg R_0$.
Our findings reveal that the exterior metric of the shell receives quantum corrections influenced by the internal mass distribution, thereby enabling an external observer to distinguish the internal structures between the shells with the same total mass.
Since the quantum corrections on the metric appear proportionally to $M_nR_n^2$, the internal shells with larger radii or heavier masses are more significant to the external observer.
Explicitly, we conducted numerical calculations for the geodesics of massive particles orbiting a particular shell configuration with uniformly distributed mass and radius across the $N$ internal shells.
In particular, focusing on shells with the empty interior and the single internal shell, we found that the quantum corrections affect the period of oscillation about the circular orbit, resulting in a larger precession angle for the case of $N=1$ as compared to the case of $N=0$.
This feature means that the presence of the internal structure amplifies the geodesic precession.
Consequently, the precession angle, which is the discrepancy between the radial and orbital periods of the quantum-corrected geodesics, increases monotonically as $N$ increases to infinity, and thus, we can identify the internal structures of stars from the quantum corrections of the metric.

In particular, in the limit $N\rightarrow\infty$, if the radius of the outermost shell in Eqs.~\eqref{eq:corrected_metric_Taylor_equidistant_Ninfty_tt} and \eqref{eq:corrected_metric_Taylor_equidistant_Ninfty_rr} is replaced by $R_0 \rightarrow \sqrt{3}R_0$, the quantum-corrected metric components exactly coincide with those in the case of $N=0$ with the radius $R_0$ of the hollow shell.
To the external observer, the quantum-corrected geodesic for the thick shell composed of an infinite number of internal shells is effectively indistinguishable from that of the hollow shell, at linear order in curvature.
Investigations at higher orders in the curvature expansion would be required to distinguish these two configurations.

We have employed the simplified assumption of identical mass and uniform distribution for each shell within a specific range.
Various mass distributions of internal shells could lead to a wide range of interesting stellar configurations, potentially revealing new regimes where quantum corrections become significant.
The issue of variations in mass density or different matter sources not only warrants further investigation but may also elucidate how internal structure affects observables including the precession angles of geodesics.

\acknowledgments
We thank Myungseok Eune and Mungon Nam for exciting discussions.
This research was supported by Basic Science Research Program through the National Research Foundation of Korea(NRF) funded by the Ministry of Education through the Center for Quantum Spacetime (CQUeST) of Sogang University (NRF-2020R1A6A1A03047877).
This work was supported by the National Research Foundation of Korea(NRF) grant funded by the Korea government(MSIT).(No. NRF-2022R1A2C1002894)


\bibliographystyle{JHEP}       

\bibliography{references}

\end{document}